\documentclass[twocolumn,aps,showpacs]{revtex4}
\usepackage{graphicx}

\input{epsf}
\begin{document}

\title{Unexpected resonant response in [Fe(001)/Cr(001)]$_{10}$ /MgO(001)
multilayers in magnetic field}
\author{F.G. Aliev, V.V.Pryadun}
\affiliation{Departamento de Fisica de la Materia Condensada, C-III, Universidad Autonoma
de Madrid, 28049, Madrid, Spain}
\author{E.Snoeck}
\affiliation{CEMES-CNRS, 29 rue Jeanne Marvig, BP 94347, 31055 Toulouse Cedex 4, France}
\date{\today}

\begin{abstract}
We observed unexpected resonant response in $[Fe/Cr]_{10}$
multilayers epitaxially grown on MgO(100) substrates which exists
only when both \emph{ac} current and \emph{dc} magnetic field are
simultaneously applied. The magnitude of the resonances is
determined by the multilayer magnetization proving their intrinsic
character. The reduction of interface epitaxy leads to non-linear
dependence of the magnitude of resonances on the alternating current
density. We speculate that the existence of the interface transition
zone could facilitate the subatomic vibrations in thin metallic
films and multilayers grown on bulk insulating substrates.
\end{abstract}

\maketitle


Two decades ago giant magnetoresistance (GMR) was discovered in
Fe/Cr multilayers \cite{Fert88, Grunberg89}. The GMR phenomenon was
found with in-plane permanent $(dc)$ electric current flowing
through the multilayer (ML) structure in the presence of $dc$
magnetic field. Recently unusual dynamic properties of metallic
multilayers such as current induced magnetization reversal
\cite{Myers99}, spin torque oscillations \cite{Russek2004},
anomalous magnetic losses at low temperatures \cite{Aliev02} were
observed. Their investigations open new horizons in the development
of new spintronic and thin film based devices.

The high density alternating current $(ac)$ flowing through a
conducting film situated in a $dc$ magnetic field creates the
longitudinal voltage drop $U_{ac}$. In addition, electromotive
resonances $U_{em}$ excited by a time dependent Lorentz force might
be observed. These resonances down to kHz frequencies could be
either due to time dependent current re-distribution in
semiconducting films \cite{ChaoticHall84} or mechanical vibrations
in suspended films \cite{clamped flexural beam,ReviewNEMS,PSS05}.
Our Letter reports on unexpected low frequency voltage excitations
in [Fe/Cr]$_{10}$ multilayers epitaxially grown on single crystal
MgO substrates. We underline that the resonances exist only with
both $ac$ current and $dc$ magnetic field \emph{simultaneously
applied}. In the low current-low field limit their magnitude varies
linearly with current and quadratically with magnetic field. We
observed qualitatively similar effects in other thin films and
multilayers such as Nb(1000$\mathring{A}$)/Si(001);
Al(30$\mathring{A}$)/Fe(150$\mathring{A}$)/MgO(001) and
Cr(200$\mathring{A}$)/MgO(001). This proves that these features are
common for different thin metallic films and multilayers grown on
insulating bulk substrates. Here we report only findings for Fe/Cr
multilayers. Firstly, because these are multilayers where GMR effect
was discovered. Secondly, Fe/Cr MLs in strong antiferromagnetic
coupling regime provide unique possibility to verify an intrinsic
character of the effects by investigating the link between magnitude
of the resonances and the sample magnetization. And finally, because
the epitaxial growth of [Fe/Cr]/MgO(100) multilayes is well
established.

The epitaxial [Fe(30$\mathring{A}$)/Cr(12$\mathring{A}$)]$_{10}$
multilayers are prepared in a molecular beam epitaxy system on
MgO(100) oriented substrates with thickness of about 1.5mm held
(unless otherwise stated) at 50 $^{0}$C and covered with an
approximately 10\AA\ thick protective Cr layer. The Cr spacer
thickness corresponds to the first antiferromagnetic peak in the
interlayer exchange coupling for the Fe/Cr system and produces a
maximum GMR of about 20\% at 300 K and 100\% at 4.2 K
\cite{Schad99}. The magnetic and electron transport characteristics
\cite{Aliev02,Aliev98} as well as TEM analysis both in conventional
and in high resolution mode (HRTEM) with point resolution of 0.12
nm. The quantitative analysis of the lattice deformation between the
substrate and the ML was carried out using the Geometrical Phase
Analysis (GPA) to study the HRTEM micrographs \cite{GPA,Snoeck}.

To measure precisely the electromotive response we employed either
five-probe scheme with in-plane magnetic field, with electric
current injected through the central contact and split in two
opposite directions or used Hall scheme with the voltage probes
situated at the opposite sides and magnetic field being
perpendicular to the film plane. The reduced $U_{ac}$ and therefore
relatively small background voltage achieved with these balanced
schemes in compare with unbalanced 4-probe method permits more clear
detection of the electromotive response $U_{em}$. Figure 1 compares
the results obtained for ML1 with 4 and 5-point schemes sketched in
the insets to Figures 1 and 2 correspondingly and by using the same
drive current and magnetic field. Only weak (about 10$^{-3}$ of the
background voltage $U_{ac})$ as demonstrates the expanded scale)
resonances have been observed with the conventional 4-probe
technique. Below we shall discuss only the measurements in the
balanced (i.e. 5-probe and Hall) geometries. Similar to the 5-probe
balanced schemes \cite{APL2002,Nature2003} were recently applied for
sensitive detection of the electromotive response in suspended
nanoelectromechanical systems. In our experiments the generated $ac$
voltage was recorded in the frequency region between few Hz to 60
kHz by using EGG-7265 lock-in amplifier (insets to Figures 1 and 2)
while the magnetic field up to 9 Tesla was created by
superconducting magnet.

In Figure 2a we compare frequency dependence of the electromotive
voltage measured by using 5-probe scheme in two [Fe/Cr]$_{10}$
MLs2,3 with different size of active (i.e. with maximum current
density) zones. Both studied structures show the presence of few
highly reproducible fundamental resonances and number of less
pronounced smaller resonances, all induced by the magnetic field. In
the sample with larger active zone we observed the appearance of new
resonances at lower frequencies (below 1kHz). The dominant
resonances should correspond to those fundamental modes whose
electric potential profiles, integrated between the voltage
contacts, have maximum values. The in-plane field rotation at 45
degrees for the ML2 (which has no influence on the resistance of the
antiferomagnetically coupled multilayer) strongly influences the
resonances (see Fig.2b). This shows importance of the direction of
in-plane magnetic field for the excited fundamental frequencies. The
inset in the Fig.1b demonstrates the evidence for quadratic
dependence of the magnitude resonances on the magnetic field in the
low field regime $(0<H<1T)$.

In order to prove further the intrinsic character of the observed
resonances in Figure 3(a-c) we compare the $ac$ and $dc$ responses.
Here we take advantage of the special feature of our synthetic Fe/Cr
antiferromagnets, which is related to field induced transition from
antiparallel to parallel (ferromagnetic) Fe layers alignment,
providing the well known giant magnetoresistance effect
\cite{Fert88, Grunberg89} (Fig.3c). Figure 3a presents the
normalized by the magnetic field $ac$ voltage response measured in
the Hall configuration for ML2 with drive frequencies close to the
main resonance ($f_{r}$ = 4210 Hz). Increasing magnetic field
broadens the resonances but weakly influences the resonant frequency
(Fig.3a). Figures 3 b,c compare the $dc$ characteristics (Hall
effect, magnetoresistance and out of plane magnetization) with
magnitude of the resonant $ac$ Hall response close to $f=f_{r}$
which shown in Fig.3a. The $dc$ Hall resistance in magnetic films is
expected to dependent on magnetization \emph{M} as:
$R_{H}=(R_{0}H+R_{A}4\pi M)$ where $R_{0}$, $R_{A}$ - ordinary and
anomalous Hall coefficients correspondingly. Figure 3b clear
demonstrates that the $dc$ and the resonant $ac$ Hall responses are
determined by the field dependence of the out-of plane magnetization
(Fig.3c) with one difference: in the resonant conditions the $ac$
Hall resistance exceeds in more than order of the magnitude the $dc$
Hall resistance. \emph{The above experiments directly prove that we
are not dealing with artifacts.}

Even higher magnitude voltage resonances (for the similar
experimental conditions) have been observed in ML4 (see Fig.4a) with
the same dimensions as ML2, but was deposited at higher temperature
(450 $^{0}$C). The multilayer ML4 has enhanced interface disorder
\cite{Schad99} which suppress the antiferromagnetic coupling and
reduces the GMR to below 20\% at low temperatures. In ML4 the main
resonances have been found to exceed notably those measured in ML2
with similar current and magnetic field allowing observation of
transition to the nonlinear resonant response. Figure 4a
demonstrates an evidence of strong deviation from the Lorenzian
shape of the resonance curves which consists in the abrupt change in
the amplitude of the response vs. frequency for the current
densities exceeding J = $0.6\times 10^{3}A/cm^{2}$.

The resonant excitations in [Fe/Cr]$_{10}$/MgO(100) multilayers are
observed up to room temperature. Figure 4b shows a typical
temperature dependence of the position and the magnitude of one of
fundamental resonances which reveals a shift in the resonance to
lower frequencies probably reflecting the thermal expansion of the
multilayer/substrate system. For metallic multilayers under study
conductance increases with decreasing temperature \cite{Aliev98},
therefore, the resonance broadening for the fixed fields with
temperature contradicts to explanation in terms of Foucault current
effect.

Our HRTEM experiments on [Fe/Cr(001)]$_{10}$ on MgO(100) (Fig.5)
show that the growth is epitaxial and no amorphous layers close to
interphase is observed. The bcc structure of Fe/Cr is rotated by 45
degree in the interface plane relative to MgO(001) substrate. The
epitaxial relation is therefore Fe/Cr(001)[110]//MgO(001)[100] as
expected for Fe/MgO(001) \cite{Schad99}. The measurements of the
relative deformation (Fig.5b) between the planes perpendicular to
the interface i.e. the (110) planes of Fe/Cr and the (200) MgO ones
is 3.3 percent. This corresponds well to the misfit between the
(002)-MgO (0.21 nm) plane and the (110)-Fe ones (0.203 nm) or
(110)-Cr (0.204 nm). We did not measure any large change of the
(110) Fe and Cr intereticular distance on the HREM images of the top
surface of the Fe/Cr) which suggest that the relaxation of the
metallic ML is constant. The Fe/Cr stacking is hardly visible in
"normal" TEM or HRTEM mode since Fe and Cr have very close atomic
number. Electron microscopy therefore show that we deal with high
quality metallic films epitaxially grown on insulating substrates.

What could be physical explanation of these resonances? Scenarios
involving different types of interface or bulk sound waves as well
as magnetic domain wall excitations imply resonant frequencies above
MHz, substantially exceeding the experimentally observed. Purely
electronic origin of resonances, similar to those previously
reported for solid state plasma in the presence of large electric
and magnetic fields \cite{ChaoticHall84} and attributed to current
filaments instabilities, is inconsistent with the observed linear
variation of the resonance magnitude with the $ac$ drive. Below we
speculate that subatomic vibrations due to interface transition zone
between substrate and multilayer might explain our experiments.

We suppose that ML is connected to MgO through interface transition
zone (ITZ) \cite{ITZ} which has a different  mechanical properties
than the rest of layers. The ITZ is originated from the lost bonds
and misfit dislocations. Indeed, Figure 5b shows clearly the
presence of regularly distributed misfit dislocations along the
interface. The rough estimation for the width of the ITZ by
evaluating spacial derivative of the stress in the direction
perpendicularly to the interface (Fig.5c) gives 2-4 nm. We believe
that notable reduction of the Young´s modulus for the ITZ \cite{ITZ}
might allow the flexural or sliding subatomic vibrations. We note
that recent studies reveal that tribology on nanoscale is very
different from the one expected for bulk materials with sliding
friction vanishing for weak transversal mechanical vibrations
\cite{SlidingResonances}.

\emph{Our hypothesis is supported by the main experimental data}
including: (i) dependence of the resonance frequency and the
magnitude on the temperature (if active zone mechanically vibrates,
fundamental frequencies are expected to decrease with temperature);
(ii) dependence on the resonance magnitude and the width on the
magnetic field as well as their dependence on the $ac$ current drive
density which are similar to those predicted and observed for the
nanoelectromechanical systems \cite{clamped flexural
beam,ReviewNEMS,PSS05,APL2002,Nature2003}. Moreover, (iii) the
nonlinear response at low current densities observed for the ML4 (Fig.4b) with suppressed GMR due to atom%
\'{}%
s interdiffusion \cite{Schad99} and the reduced (in comparison with
MLs1-3) epitaxy of the ITZ is also along with our hypothesis.
Finally, (iv) the quality factors observed Q $\sim 10^{3}-10^{4}$
are close to those expected from the phenomenological $Q$ vs. volume
scaling reported for Si- based nanoelectromechanical systems
\cite{ReviewNEMS}.

Within the proposed model, the electromotive voltage generated by
the vibrations depends on the modes excited and the contacts
configuration. Assuming low damping \cite{clamped flexural beam},
the voltage generated by flexural resonance is : $U_{em}\approx \xi
QB^{2}L^{2}I/2\pi mf$, where $\xi $ is a constant of the order of
unity which depends on the shape of resonant mode, $m$ - mass, $L$ -
conductance length and $Q$- quality factor. For $Q\approx 10^{3}$;
$B$ =1Tesla, $L=100\mu m $, $m=10^{-12}$ kg, $f\approx 10^{4}$ Hz
and current $I=10^{-4}A$ we get electromotive voltage $U_{em}\approx
2.5\ast 10^{-4}V$ which exceeds in about one order the
experimentally observed values.

The above model also explains typical values of the observed
resonance frequencies. By using Young%
s values $E\approx 10^{11}$ Pa typically reported for the metallic
multilayers \cite{YoungsModMML}, mass density and geometrical factors of the
actively vibrating zones of the patterned $[Fe/Cr]_{10}$ multilayers ($L_{0}$%
=300-100 $\mu $m), we obtain the fundamental flexural frequencies
$f_{0}\sim 1.5-15$ kHz close to the experimentally measured. Arrows
in Figure 1a indicate relative (normalized to the lowest in
frequency mode with largest amplitude for ML2) position of the three
main resonances expected within the above model in the conditions
when only flexural resonances are excited by in plane magnetic field
perpendicular to the $ac$ current. Having in mind contact attached
to the central part, we find agreement with the model to be
satisfactory.

In \emph{Conclusions}, we have observed unexpected intrinsic low
frequency resonant response in Fe/Cr multilayers only presented with
both \emph{ac} current and \emph{dc} magnetic field applied.
Although the physical mechanism behind still remains to be fully
clarified, the consistent experimental data and simple estimations
support the possibility of the excitation of subatomic vibrations in
metallic thin films epitaxially grown on bulk substrates.

The authors acknowledge R.Schad for samples preparation, T.Alieva
and A.Levanuyk for fruitful discussions.

\section*{FIGURE CAPTIONS}

\textbf{FIG.1}\newline Color online: Comparison of the frequiency
dependent response in the 4 and 5-probe schemes for ML1 (7.5 by 30
$\mu$m active zone) measured at 10K with in-plane magnetic field of
3T perpendicular to the current of 25$\mu$A. The inset explains the
4-probe measurement scheme.

\textbf{FIG.2}\newline Color online: (a) Frequency dependence of the
$U_{em}$ measured at 6K for the in plane magnetic field of 1T for
two [Fe/Cr]$_{10} $ on MgO multilayers with different dimensions.
Current of 100$\mu$A (ML2) and of 10$\mu$A (ML3) was used to provide
the same current density of about $2.5 \times 10 ^{3}$ A/cm$^{2}$.
The inset sketches the 5-probe measurement scheme. Vertical arrows
are described in the text. (b) Frequency dependence of the $U_{em}$
measured for the ML2 in the same conditions as in the part(a) but
with the in-plane fields of 1T and of 0.07T rotated at $45^0$. The
inset plots dependence of the magnitude of one resonances on the
square of the magnetic field.

\textbf{FIG.3}\newline Color online: (a) Magnitude of one of the
dominant resonances normalized by the magnetic field and measured
for ML2 at T = 10K in the Hall geometry with 50$\mu$A (current
density of about $1.25 \times 10^{3} A/cm^{2}$). (b) Comparison
between field dependent $dc$ Hall resistance and resonant $ac$ Hall
resistance ($f_{r} = 4210 Hz$, divided by factor of 30) measured for
T = 10 K. (c) Field dependences of the out of plane
magnetoresistance and of the magnetization for MML2 measured at 10K.

\textbf{FIG.4}\newline Color online: (a) Normalised by the applied
current of 2,10,25 and 100$\mu$A electromotive response close to the
dominant resonance in ML4 measured with H=1T and with the 5-probe
scheme, similarly to shown in Fig.1. (b) Temperature dependence of
one of the dominant resonances measured with the 5-probe
configuration for the ML2 with H=1Tesla and with current of
100$\mu$A ($2.5 \times 10^{3} A/cm^{2}$). The inset shows
temperature dependence of the resonant frequency and of the
corresponding Q-factor.

\textbf{FIG.5}\newline Color online: (a) HRTEM micrograph of the
MgO/[Fe/Cr] interface studied along the [100]-MgO zone axis. (b)
$\varepsilon_{xx}$ deformation image calculated from (a) (with "x"
being parallel to the interface). Note the regular distribution of
misfit dislocations (arrows) along the interface. (c) Typical
variation of the relative deformation across the interface between
Fe/Cr multilayer and MgO substrate.

\newpage

\begin{figure}[tbp]
\par
\begin{center}
\includegraphics[width=7.5cm]{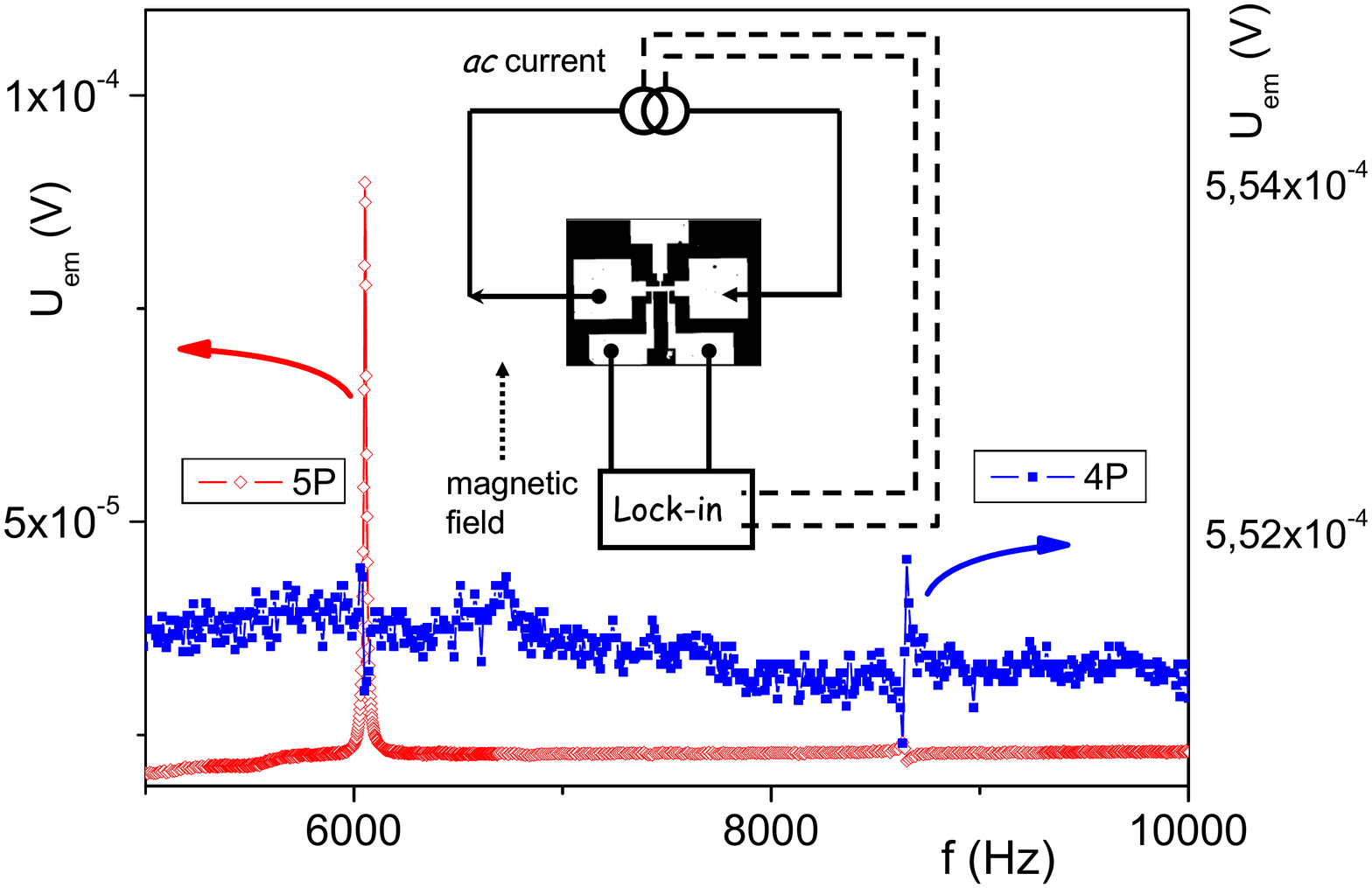}
\end{center}
\caption{}
\label{Fig1}
\end{figure}

\begin{figure}[tbp]
\par
\begin{center}
\includegraphics[width=7.5cm]{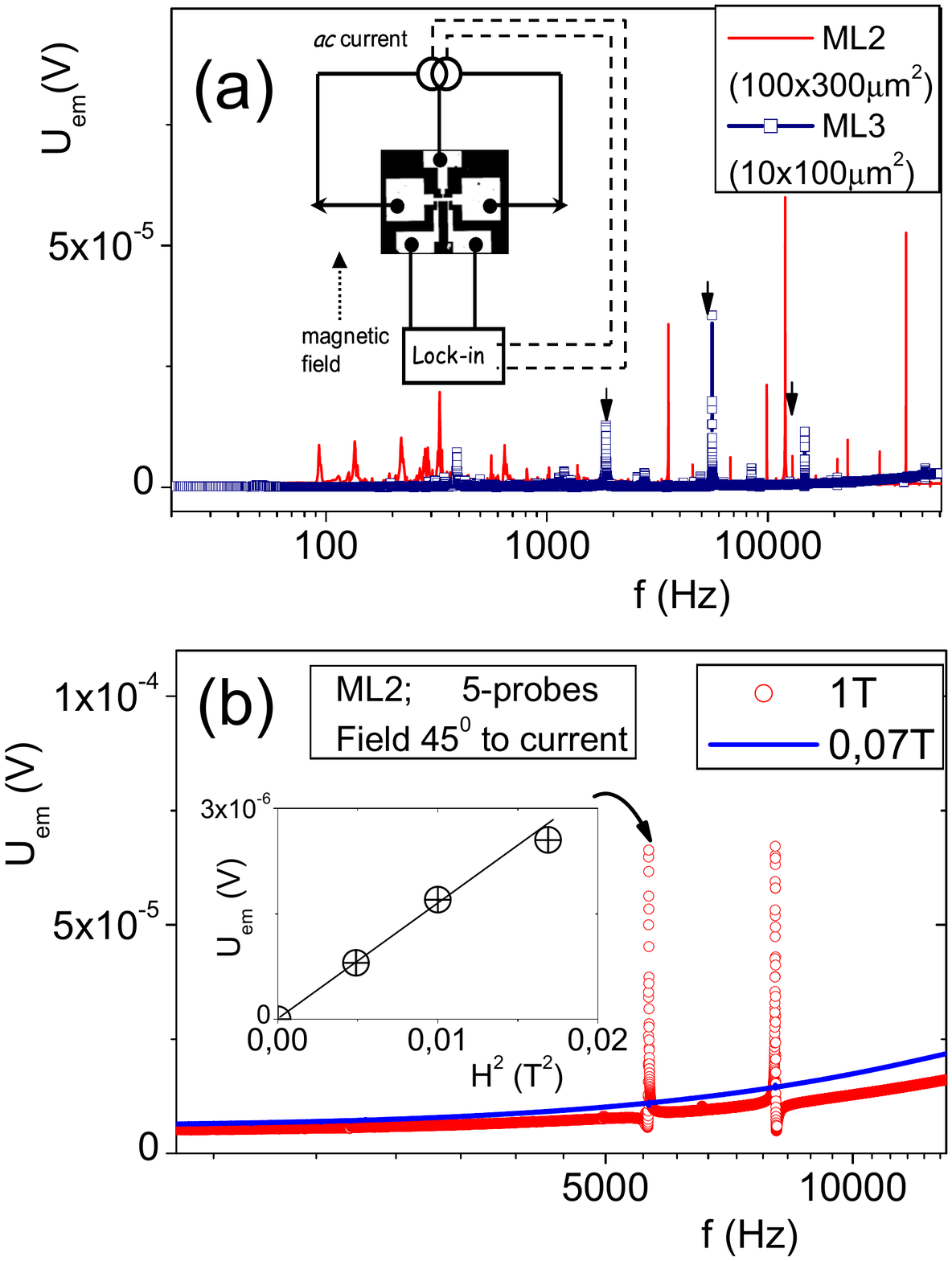}
\end{center}
\caption{} \label{Fig2}
\end{figure}

\begin{figure}[tbp]
\par
\begin{center}
\includegraphics[width=7.5cm]{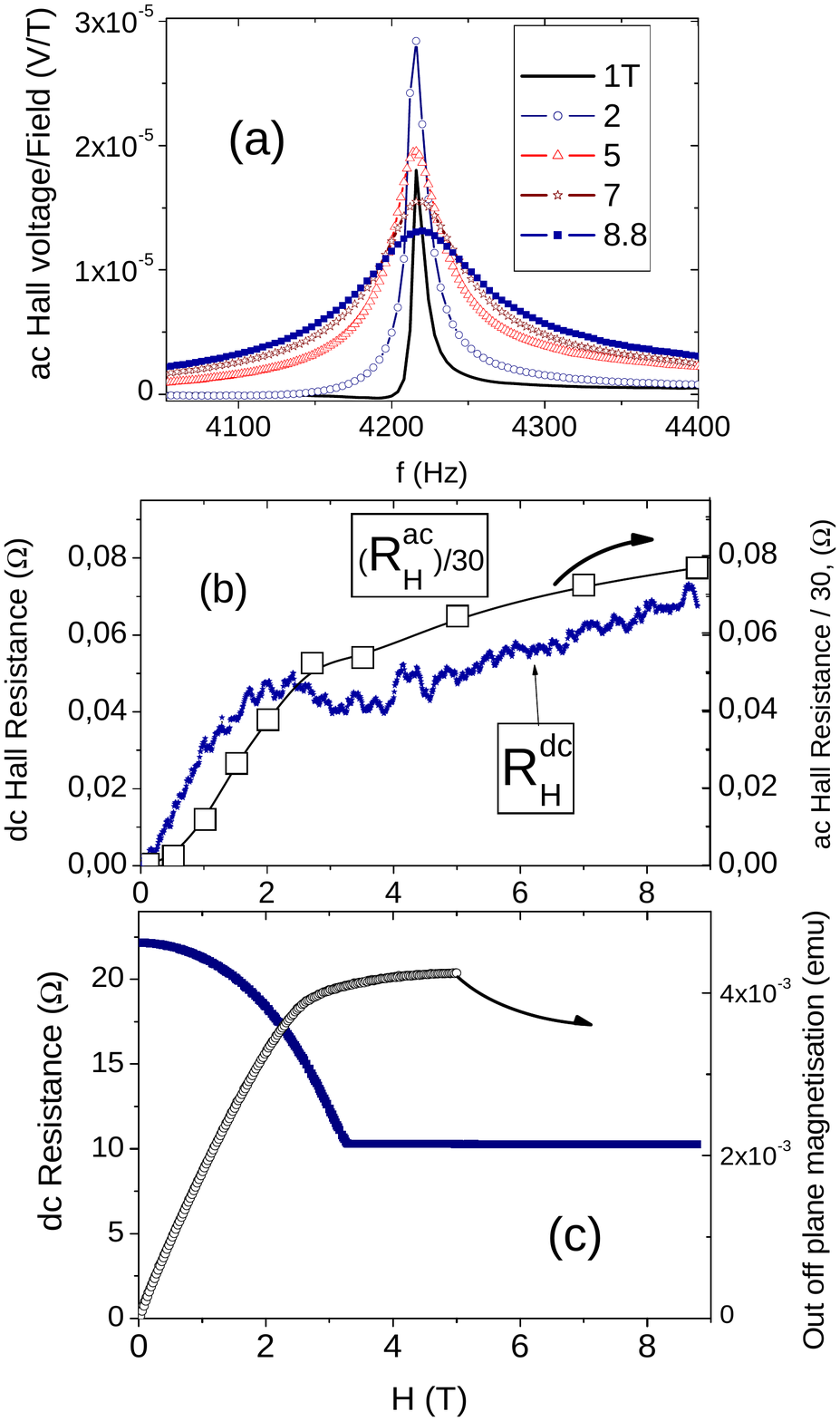}
\end{center}
\caption{} \label{Fig3}
\end{figure}

\begin{figure}[tbp]
\par
\begin{center}
\includegraphics[width=7.5cm]{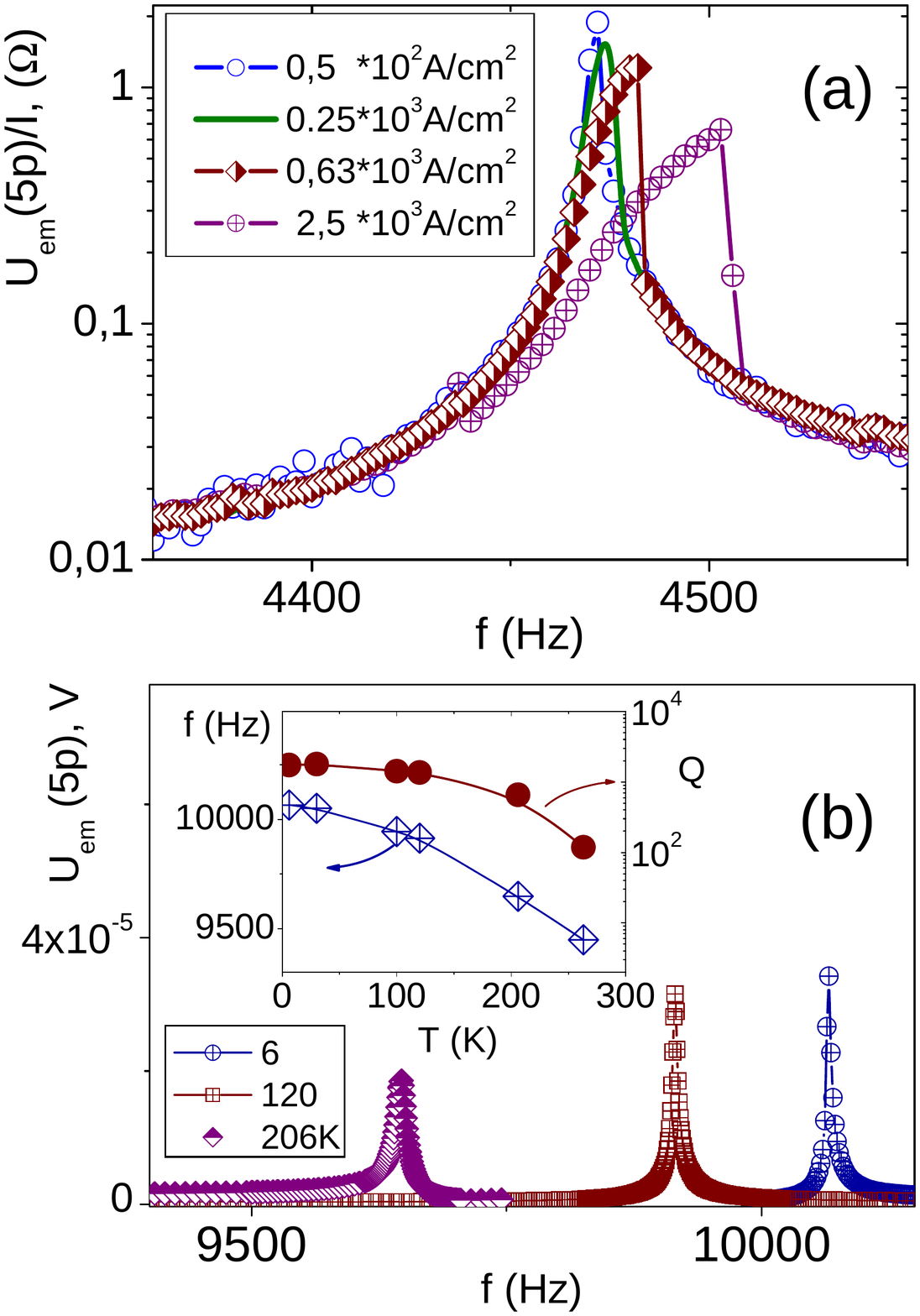}
\end{center}
\caption{} \label{Fig4}
\end{figure}

\begin{figure}[tbp]
\par
\begin{center}
\includegraphics[width=7.5cm]{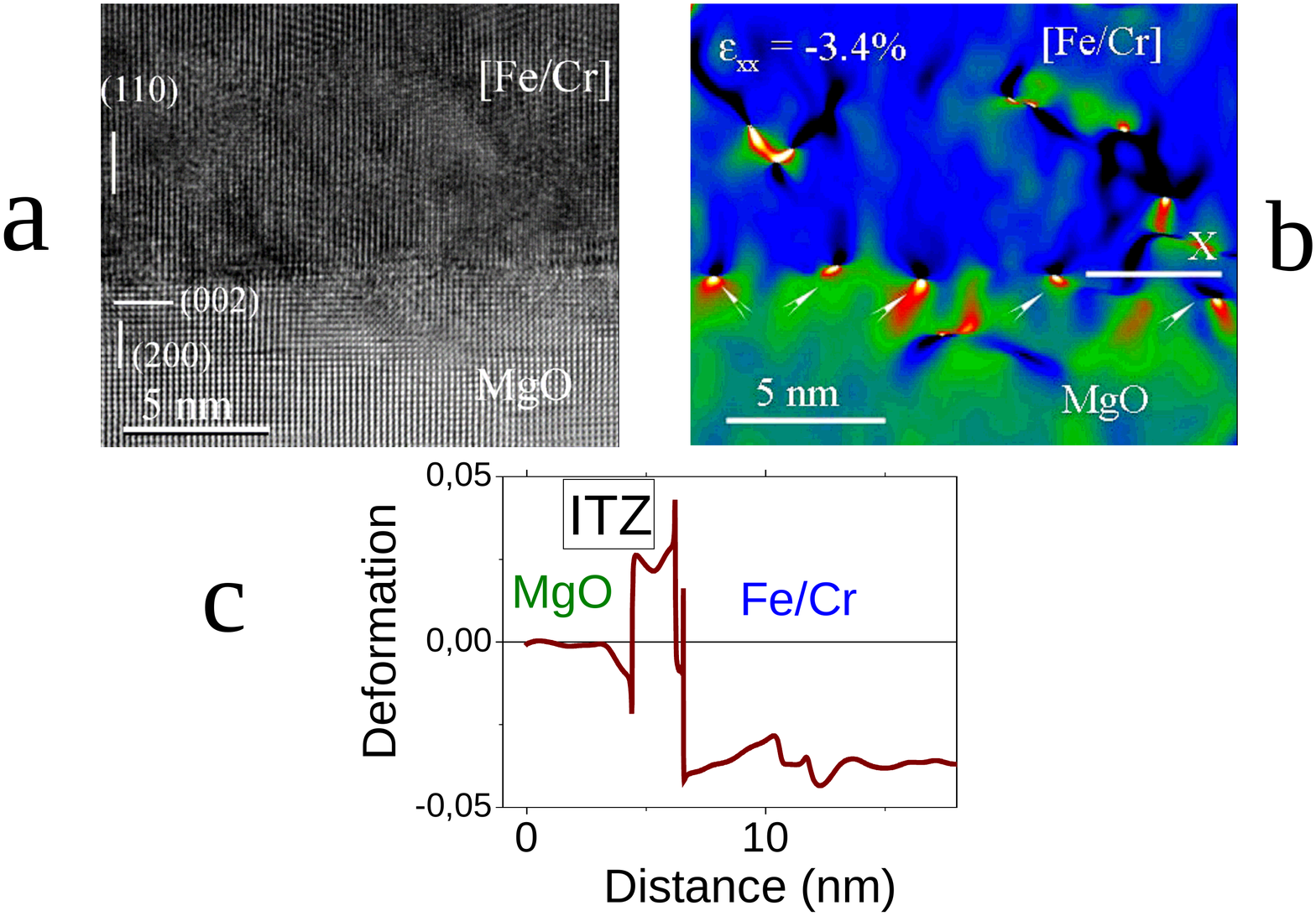}
\end{center}
\caption{}
\label{Fig5}
\end{figure}

\end{document}